\documentclass[preprint,12pt]{elsarticle}
\journal{Annals of Physics}
\RequirePackage{bm,,amsfonts,lineno,hyperref,amsmath,amssymb,microtype,float}

\begin{document}

\begin{frontmatter}

\title{Relativistic model for anisotropic strange stars}

\author{Debabrata Deb}
\address{Department of Physics, Indian Institute of Engineering Science 
	and Technology, Shibpur, Howrah, West Bengal, 711103, India\\ddeb.rs2016@physics.iiests.ac.in}

\author{Sourav Roy Chowdhury}
\address{Department of Physics, Indian Institute of Engineering Science 
	and Technology, Shibpur, Howrah, West Bengal, 711103, India\\sourav.rs2016@physics.iiests.ac.in}

\author{Saibal Ray}
\address{Department of Physics, Government College of Engineering and
Ceramic Technology, Kolkata 700010, West Bengal, India\\saibal@associates.iucaa.in}

\author{Farook Rahaman}
\address{Department of Mathematics, Jadavpur University, Kolkata 700032, West Bengal, India\\rahaman@associates.iucaa.in}

\author{B.K. Guha}
 \address{Department of Physics, Indian Institute of Engineering Science 
	and Technology, Shibpur, Howrah, West Bengal, 711103, India\\bkg@physics.iiests.ac.in}

\date{Received: date / Accepted: date}

\maketitle

\begin{abstract}
	In this article, we attempt to find a singularity free solution of Einstein's field equations for 
    compact stellar objects, precisely strange (quark) stars, considering Schwarzschild metric as the 
	exterior spacetime. To this end, we consider that the stellar object is spherically 
    symmetric, static and anisotropic in nature and follows the density profile given by Mak and 
    Harko (2002), which satisfies all the physical conditions. 
    To investigate different properties of the ultra-dense strange stars we have employed the MIT bag model 
    for the quark matter. Our investigation displays an interesting feature that the anisotropy of compact stars  
    increases with the radial coordinate and attains its maximum value at the surface which seems an inherent property for 
    the singularity free anisotropic compact stellar objects. In this connection we also perform several tests for physical
	features of the proposed model and show that these are reasonably acceptable within certain range. Further, we 
	find that the model is consistent with the energy conditions and the compact stellar structure is stable with the 
	validity of the TOV equation and Herrera cracking concept. For the masses bellow the maximum mass point in 
	mass vs radius curve the typical behavior achieved within the framework of general relativity. We have calculated 
    the maximum mass and radius of the strange stars for the three finite values of bag constant $B_g$.   
\end{abstract}

\begin{keyword}
General Relativity; equation of state; exact solution; Strange stars
\end{keyword}

\end{frontmatter}

\section{Introduction}\label{sec1}  
Anisotropy is the property of being directionally dependent, which in the present context 
can be defined as a difference of pressures ($P_t -P_r$, the tangential and radial pressures 
respectively for the fluid sphere), may arise due to different causes. 
In relativistic astrophysics it has been suggested that anisotropy arises due to
presence of mixture of fluids, rotational motion, existence of superfluid,
presence of magnetic field or external field, phase transition so on. In early stages of 
research on astrophysical anisotropy Ruderman \cite{Ruderman1972} especially conceived 
that for the high density anisotropy is the inherent
property of nuclear matters where interactions are relativistic in nature. 
On the other hand, Bowers and Liang \cite{BL1974} argued that
superdense matter may be anisotropic due to superfluidity and 
superconductivity in the presence of complex strong interactions. 
A detailed review work is available by Herrera and Santos \cite{HS1997} where
they have discussed possible causes for the appearance of local anisotropy 
(principal stresses unequal) in selfgravitating systems and presented its main consequences. 
However, in the later stage diversified investigations to understand the nature and
consequences have been performed by several scientists on ultra-dense spherically symmetric fluid
spheres having pressure anisotropy~\cite{Ivanov2002,SM2003,MH2003,Usov2004,Varela2010,Rahaman2010,Rahaman2011,Rahaman2012,Kalam2012,Deb2015,Shee2016,Maurya2016,Maurya2017}.

However, to overcome any misunderstanding it is to note in the initial stage that the present model being static one we have ignored the magnetic field associated with the star and hence, anisotropy has not been introduced due to affect of the magnetic field. In the model it is assumed that solely anisotropic stress generated here due to the anisotropic fluid or the elastic nature of the super fluid or the phase transition. Several investigators already have published papers with different reasons of the anisotropic stress, except rotational effect or magnetic effect~\cite{BL1974,MH2002,Dev2003,Kalam2005,Horvat2011}.

In the context of the physical properties and internal structure of
the compact stars Itoh~\cite{Itoh1970} first
suggested that the quark stars may exist in hydrostatic equilibrium. According
to Bodmer~\cite{Bodmer1971} the quark matter made of $u$, $d$ and $s$
quarks is more stable than any ordinary nuclear matter.
Cheng et al.~\cite{Cheng1998} argued that after a supernova
explosion, a massive star's core collapses to a strange star.
It was also proposed by Alford~\cite{Alford2001} that in the dense core
of a neutron star there is sufficiently high density and low temperature
to crush the hadrons into quark matter.

It is quite interesting that the equation of state (EOS) chosen by the following authors in Refs.~\cite{Alcock1986,Haensel1986,Weber2005,Perez2010,Rodrigues2011,Bordbar2011}
for neutron stars did not explain the properties of the observed compact stars,
like $PSR~J1614-2230$, $Vela~X-1$, $PSR~J1903+327$, $Cen~X-3$, $LMC~X-4$,
whereas the strange matter EOS can explain them. In the interior of such
stars there may exist many exotic phases. It is observed that to describe a stellar model 
made of strange quark matter well known MIT bag EOS can be considered under such a situation. 
Successful use of the MIT bag EOS to describe strange quark stars the following 
recent works are available in the literature~\cite{1,2,3,4,5,6,7,8}. This type of simplest form 
of EOS is very useful to study equilibrium configuration of a compact stellar object made of 
only up, down and strange quarks even in the framework of general relativity without invoking
any quantum mechanical particle physics aspect. 

Hence, in the present work our investigation will be mostly restricted to the EOS of 
strange matter only. It is to mention here that Rahaman et al.~\cite{Rahaman2014} 
by using the MIT bag model proposed a model for strange stars where they 
were able to analyze the physical properties of the stars from 6 km to the surface.
Therefore, here our aim will be to choose a phenomenological MIT bag model~\cite{Chodos1974,Farhi1984,Witten1984} 
where the vacuum pressure $B_g$ confirms the quark confinement that equilibrates 
the pressure of quarks to stabilize the system. We have proposed a new stable model 
for strange stars which is neutral, static as well as anisotropic fluid sphere 
valid under the appropriate physical condition. Using the observed values of the mass of strange stars 
we have analyzed their physical properties and also predicted respective radii of the stars. 

The above study is organized as follows: in Sec. 2 and Sec. 3 basic equations
and the appropriate EOS are respectively provided. We have sought for solutions in Sec. 4 
whereas various physical features of the strange stars are discussed in Sec. 5. The article
is concluded with a comparative study using provided data sets for the different
strange stars and a short discussion in Sec. 6.

\section{Basic equations of stellar structure}\label{sec2}
We consider that the interior of the strange stars is well
described by the following space time metric
\begin{equation}
{ds}^{2}=-{{e}^{\nu(r)}}{{dt}^{2}}+{{e}^{\lambda(r)}}{{dr}^{2}}+{r}^{2}({{d\theta}^{2}}+{{sin}^{2}}\theta{{d\phi}^{2}}),\label{eq2}
\end{equation}
where $\nu$ and $\lambda$ are functions of the radial coordinate $r$ only.

The energy-momentum tensor for the proposed strange star model is given by
\begin{eqnarray}\label{eq3}
&\qquad\hspace{2cm} T^{0}_{0}=\rho,\\ \label{eq4}
&\qquad\hspace{2cm} T^{1}_{1}=-P_r,\\ \label{eq5}
&\qquad\hspace{2cm} T^{2}_{2}=T^{3}_{3}=-P_t,
\end{eqnarray}
where $\rho$, $P_r$ and $P_t$ in Eqs. (\ref{eq3}), (\ref{eq4}) and (\ref{eq5})
represent the energy density, radial and tangential pressures respectively for the fluid sphere.

The Einstein field equations are now can be listed below
\begin{eqnarray}\label{eq6}
&\qquad\hspace{-0.2cm} {{\rm e}^{-\lambda}} \left( {\frac {{\lambda}^{\prime}}{r}}-\frac{1}{r^{2}} \right) +{\frac{1}{{r}^{2}}}=8\,\pi\,
\rho,\\ \label{eq7}
&\qquad\hspace{-0.2cm} {{\rm e}^{-\lambda}}\left({\frac {1}{r^{2}}}+\frac{{\nu}^{\prime}}{r} \right)-{\frac{1}{{r}^{2}}}=8\,\pi\,{P_r},\\ \label{eq8}
&\qquad\hspace{-0.2cm} \frac{1}{2}{{\rm e}^{-\lambda}}\left[\frac{1}{2}{\left({\nu}^{\prime}\right)}^{2}+{\nu}^{\prime\prime}-
\frac{1}{2}{\lambda}^{\prime}{\nu}^{\prime}+\frac{1}{r}\left({\nu}^{\prime}-{\lambda}^{\prime}\right)\right]=8\,\pi\,{P_t}. \nonumber \\
\end{eqnarray}

Let us define the mass function $m(r)$ of the star as
\begin{equation}
m \left( r \right) =4\,\pi\,\int_{0}^{r}\!\rho \left( r \right) {r}^
{2}{dr}.\label{eq9}
\end{equation}

Now assuming that the matter density profile inside the quark stars
can be described following Mak and Harko~\cite{MH2002} as
\begin{equation}
\rho(r)=\rho_c\left[1-\left(1-\frac{\rho_0}{\rho_c}\right)\frac{r^{2}}{R^{2}}\right],\label{eq10}
\end{equation}
where $\rho_c$ and $\rho_0$ are the central and surface densities of the star
of radius $R$ respectively.

Hence using Eqs. (\ref{eq9}) and (\ref{eq10}) we find the total mass of the quark star as follows:
\begin{equation}
M={\frac {4}{15}}\pi \, \left( 2\,\rho_{{c}}+3\,\rho_{{0}} \right) {R}^{3}. \label{eq11}
\end{equation}

\section{Equation of State (EOS)}\label{sec2.1}
The simplest form of EOS we have taken into account here is that all the three flavors 
of quarks are non-interacting and confined in a bag which can be expressed as follows
\begin{equation}
{P_r}+{B_g}={\sum_f}{P^f},\label{2.1.1}
\end{equation}
where the individual pressure $P^f$ (flavor, $f= u, d, s$ of up, down and strange quarks) 
due to all three quarks are counterbalanced by the total external bag pressure $B_g$ 
which is a constant quantity within a numerical range. 

The total energy density $\rho$ of the de-confined quarks inside the bag is given as
\begin{equation}
\rho={\sum_f}{{\rho}^f}+B_g, \label{2.1.2}
\end{equation}
where ${{\rho}^f}=3\,P_f$ is energy density of the individual quarks. 

From Eqs. (\ref{2.1.1}) and (\ref{2.1.2}) the EOS of the matter distribution can be given as
\begin{equation}
P_r=\frac{1}{3}(\rho-4\,B_g).\label{eq1}
\end{equation} 

As the radial pressure must be zero at the surface, so from Eq. (\ref{eq1}) we have
\begin{equation}
\rho_0=4\,B_g, \label{eq11a}
\end{equation}
where $\rho_0$ is the surface density at the surface $r=R$.

In our study of strange stars we have considered values of bag constant as 83~$MeV/{{(fm)}^{3}}$~\cite{Rahaman2014}, 100~$MeV/{{(fm)}^{3}}$ and 120 $MeV/{{(fm)}^{3}}$.

\section{Solutions of the basic equations}\label{sec2.2}
Now with the choice of EOS (\ref{eq1}) and using Eqs. (\ref{eq2}),
(\ref{eq6}) - (\ref{eq10}) and (\ref{eq11a}) we obtain the following physical parameters:
\begin{eqnarray}\label{eq12}
&\qquad\hspace{-1cm} \lambda \left( r \right) =-\ln  \left[ 1-\frac{8}{3}\,\pi \,{r}^{2}\rho_{{c}}+
\frac{8}{5}\,{\frac {\pi \,{r}^{4}}{{R}^
{2}}}\left( \rho_{{c}}-\rho_{{0}} \right) \right],\\ \label{eq13}
&\qquad\hspace{-1cm} \nu \left( r \right)=\Bigg[\frac{B}{A} \left\lbrace {\it arctanh} \left( {\frac {C}{A}} \right) -{\it arctanh} \left( {\frac {D}{A}} \right)  \right\rbrace -\frac{2}{3}\ln  \left( E{r}^{4}+F \right) + G\Bigg], 
 \\\label{eq14}
&\qquad P_r=\frac{1}{3}\left(\rho_c-\rho_0\right)\left[1-\frac{r^{2}}{R^{2}}\right],\\ \label{eq15}
&\qquad P_t=\frac{{c_1}({c_2}{r}^{4}-{c_3})-16\pi{{c_1}^{2}}{{r}^{6}}-10{c_4}{r}^{2}}{120{R}^{2}\left[\pi{{r}^{2}}{\rho_c}{R}^{2}
	-\frac{3}{5}{c_1}\pi{{r}^{4}}-\frac{3}{8}{R}^{2}\right]},\\ \label{eq16}
&\qquad {\Delta(r)}={P_t}-{P_r}={\frac {[ \left( 12\,{c_{{1}}}^{2}{r}^{4}-\frac{1}{5}\,c_{{1}}c_{{5}}{r}^{2}
		+c_{{6}} \right) \pi -\frac{9}{2}\,c_{{1}}{R}^{2}]{r}^{2}}{36{R}^{2}[{r}^{2} \left( \frac{3}{5}\,c_{{1}}{r}^{2}
		-\rho_{{c}}{R}^{2} \right) \pi +\frac{3}{8}\,{R}^{2}]}},
\end{eqnarray}
where
$A=\sqrt {{\pi}^{2}{R}^{2} \left[10\,\pi \,{\rho_{{c}}}^{2}{R}^{2}-9(\rho_c+\rho_0)\right] }$,
$B= \sqrt {10} \pi \,{R}^{2} \left( \rho_{{0}}-2/3\,\rho_{{c}} \right)$,
$C= \sqrt {10} \frac{\pi}{5}{R}^{2} \left( -\rho_{{c}}+6\,\rho_{{0}} \right)$,
$D= \sqrt {10} \frac{\pi}{5} \left( -6\,{r}^{2}\rho_{{c}}+6\,{r}^{2}\rho_{{0}}+5\,\rho_
{{c}}{R}^{2} \right)$,
$E= \pi \left( -24\,\rho_{{c}}+24\,\rho_{{0}} \right)$,
$F= \pi (40\,\,{r}^{2}\rho_{{c}}-15)R^2$,
$G=\frac{2}{3}\ln  \left[ -15\,{R}^{2}+ \left( 16\,\rho_{{c}}+24\,\rho_{{0}}
\right) \pi \,{R}^{4} \right]-{\frac {16}{15}}\,{R}^{2} \left( \frac{2}{3}
\rho_{{c}}+\rho_{{0}} \right) \pi +\frac{2}{3}$,
${c_1}=({\rho_c}-{\rho_0})$,~${c_2}=32(\frac{5}{4}{\rho_c}-{\rho_0})\pi{R}^{4}$,~${c_3}=15{R}^{4}$,
${c_4}=10\left[\pi(4{\rho_c}^{2}-2{{\rho}_{c}}{\rho_0}+{\rho_0}^{2}){R}^{2}-3({\rho_c}-{\rho_0})\right]{{R}^{2}}$,
${c_5}={\frac{1}{5}}\left(156\,\rho_{{c}}-84\,\rho_{0}\right){R}^{2}$,
${c_6}=\left(6\,\rho_{{c}}-3\,\rho_{0}\right)\left(4\,\rho_{{c}}-\rho_{{0}}\right){R}^{4}$ and the constant density $\rho_c$ is the central density at the centre ($r=0$).


\begin{figure}[!htp]\centering
	\includegraphics[scale=.3]{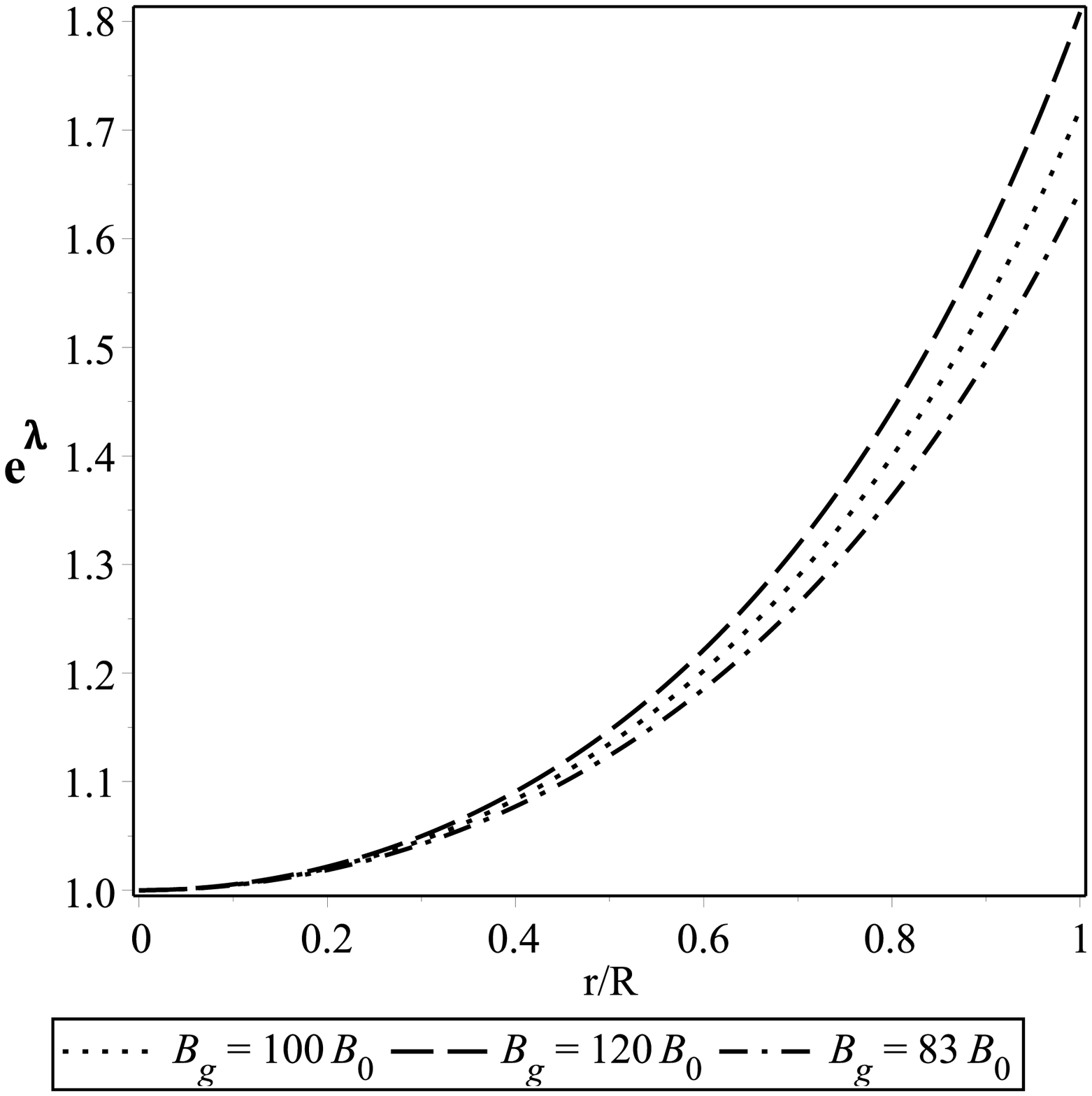}
	\includegraphics[scale=.3]{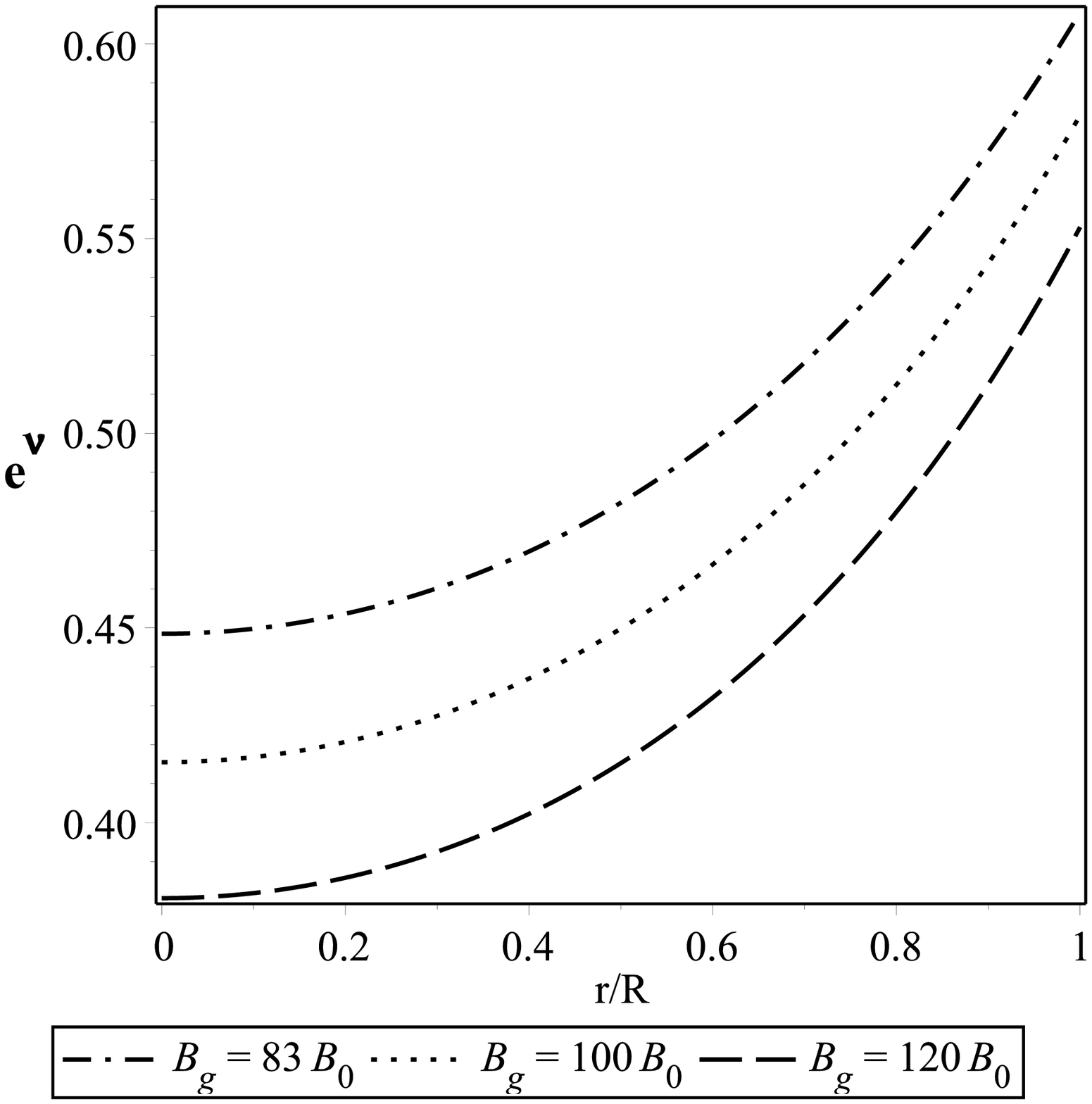}
	\caption{Variation of (i) ${e}^{\lambda(r)}$ (on the left panel) and (ii) ${e}^{\nu(r)}$
(on the right panel) as a function of the fractional radial coordinate $r/R$ for the strange star $LMC~X-4$. Here $B_0$ denotes $1 MeV/{{fm}^3}$} \label{Fig1}
\end{figure}



\begin{figure}[!htp]\centering
\includegraphics[scale=.3]{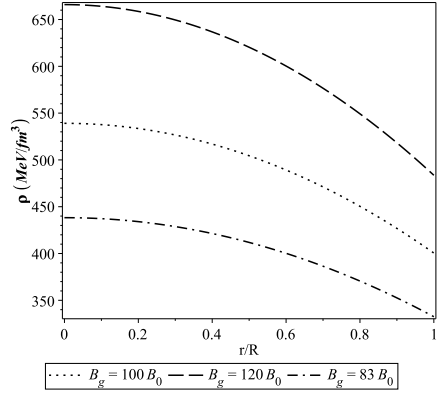}
\caption{Variation of the density $(\rho)$ as a function of the fractional radial coordinate $r/R$ for the strange star $LMC~X-4$. Here $B_0$ denotes $1~MeV/{fm^3}$} \label{Fig2}
\end{figure}



\begin{figure}[!htp]\centering
\includegraphics[scale=.3]{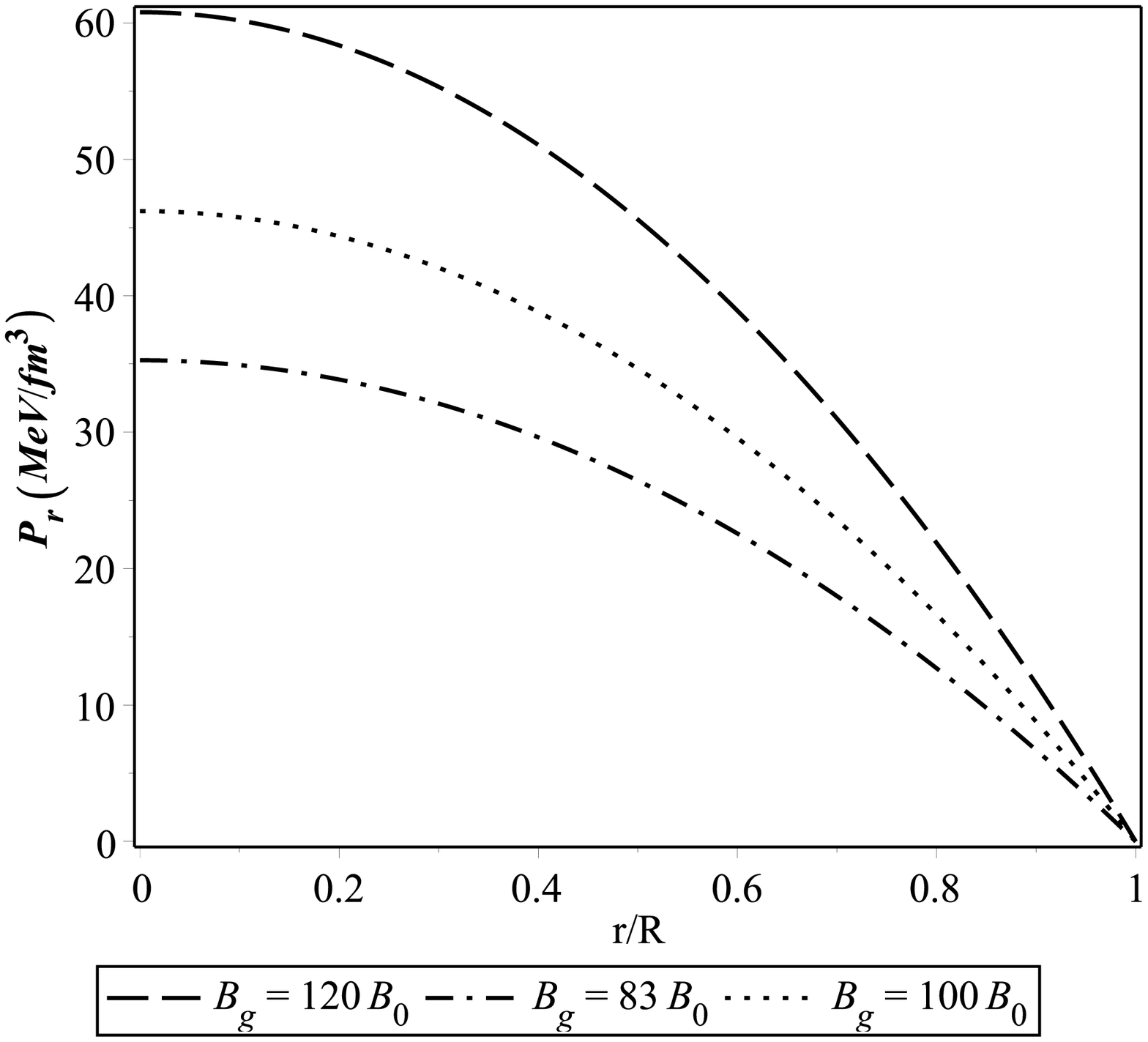}
\includegraphics[scale=.3]{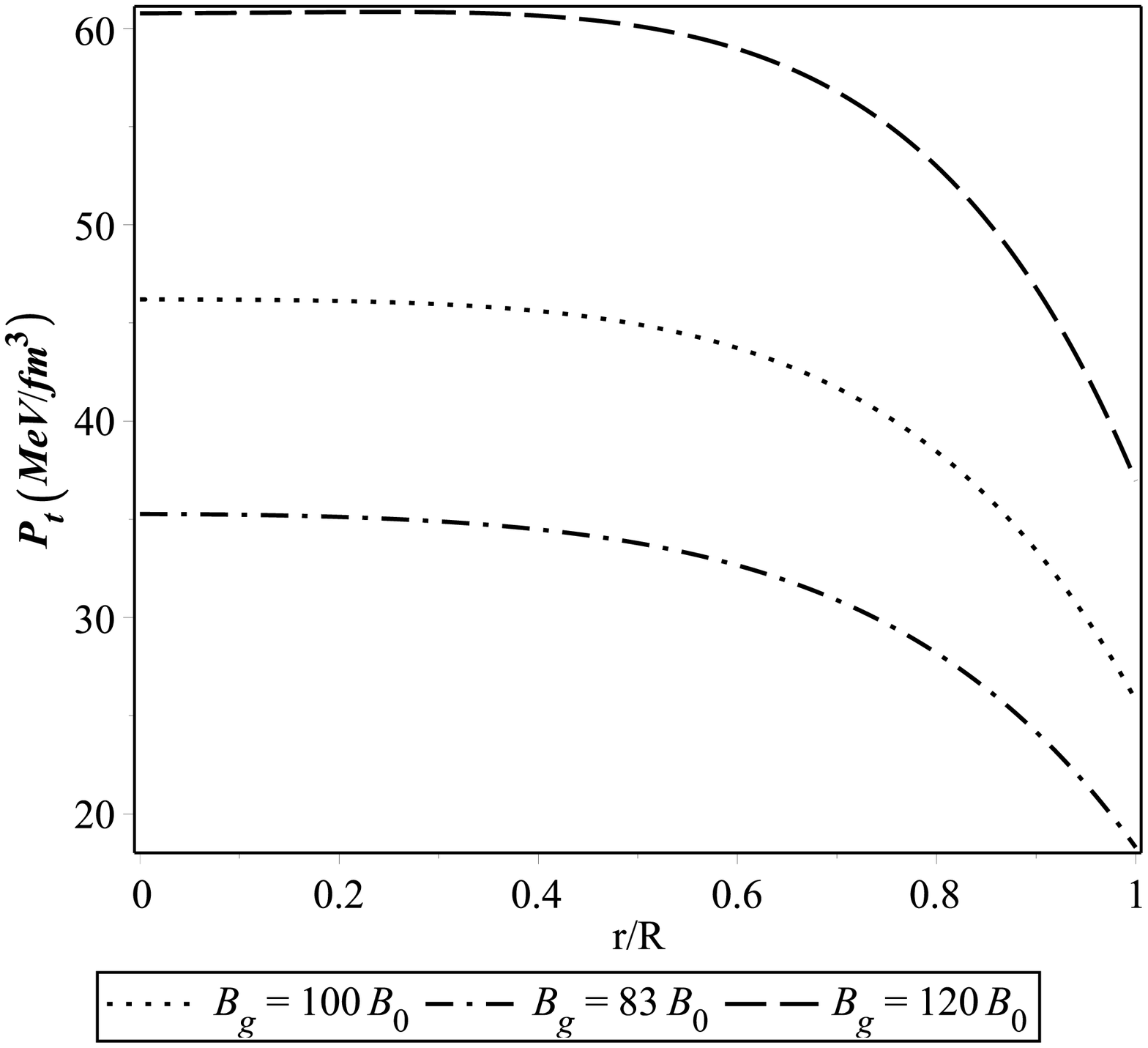}
\caption{Variation of the radial pressure, $P_r$ (on the left panel) and transverse pressure, $P_t$ (on the right panel) as a function
		of the fractional radial coordinate $r/R$ for the strange star $LMC~X-4$.
	Here $B_0$ denotes $1~MeV/{fm^3}$} \label{Fig3}
\end{figure}



\begin{figure}[!htp]\centering
\includegraphics[scale=.3]{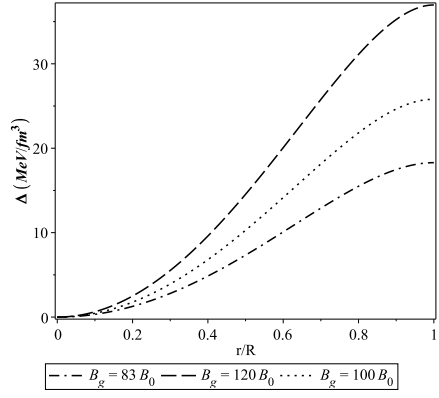}
\caption{Variation of anisotropy $(\Delta)$ as a function of the fractional radial coordinate $r/R$
	for the strange star $LMC~X-4$.	Here $B_0$ denotes $1~MeV/{fm^3}$} \label{Fig4}
\end{figure}


In Eq.~(13) we have shown the simplest form of EOS of the quark matter. This can be considered as radial part of the EOS which consist of the effect of radial pressure in the form $P_r ={\omega}_r{\rho}$. To consider the effect of the tangential pressure, the EOS can be shown as $P_t={\omega}_t{\rho}$, where ${\omega}_r$ and ${\omega}_t$ are respectively the EOS parameters which can be evaluated as
\begin{equation}
\omega_{{r}}={\frac { \left( \rho_{{c}}-4\,B_{{g}} \right)  \left( {R}
^{2}-{r}^{2} \right) }{ \left( 12\,B_{{g}}-3\,\rho_{{c}} \right) {r}^{
2}+3\,\rho_{{c}}{R}^{2}}},
\end{equation}

\begin{eqnarray}
\omega_{{t}}= \Big[-16\,\pi \, \left( -\rho_{{c}}+4\,B_{{g}}
 \right) ^{2}{r}^{6}+32\, \left( 4\,B_{{g}}-\frac{5}{4}\rho_{{c}} \right) 
 \left( -\rho_{{c}}+4\,B_{{g}} \right) {R}^{2}\pi \,{r}^{4}\nonumber \\
 -10\, \lbrace \pi \, \left( 16\,{B_{{g}}}^{2}-8\,B_{{g}}\rho_{{c}}+4\,{\rho_
{{c}}}^{2} \right) {R}^{2}+12\,B_{{g}}-3\,\rho_{{c}} \rbrace {R}^{2}{r
}^{2}+15\,{R}^{4} \left( -\rho_{{c}}+4\,B_{{g}} \right) \Big]\Bigg/\nonumber \\
\Big[ \lbrace 120
\,\rho_{{c}}{R}^{2}+120\, \left( -\rho_{{c}}+4\,B_{{g}} \right) {r}^{2
} \rbrace  \lbrace \frac{3}{5}\,\pi \, \left( -\rho_{{c}}+4\,B_{{g}} \right) {r
}^{4}+\pi \,{r}^{2}\rho_{{c}}{R}^{2}-\frac{3}{8}{R}^{2} \rbrace \Big].
\end{eqnarray}


\begin{figure}[!htp]\centering
\includegraphics[scale=.3]{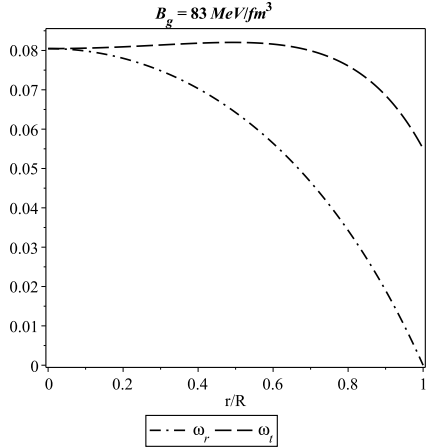}
\caption{Variation of the radial (${\omega}_r$) and tangential (${\omega}_t$) EOS parameters as a function of the fractional radial coordinate $r/R$ for the strange star $LMC~X-4$} \label{Fig5}
\end{figure}


The behaviors of the above physical parameters are shown in Figs. \ref{Fig1} - \ref{Fig5}
which are quite satisfactory.

Now let us maximize the anisotropic stress at the surface ($r=R$) without presumption
of the nature of extremum. Thus from Eq. (\ref{eq16}) we obtain:
\begin{eqnarray}
&\qquad\hspace{-1cm}\rho_c=-\Bigg[{\frac{-28\pi {{R}^{2}}{{\rho}_0}+16 {{R}^{4}}{\pi}^{2}{{{\rho}_0}}^{2}
		-15}{32\,{R}^{2}\pi \, \left( 2\,\pi \,{R}^{2}\rho_{{0}}+1 \right) }}\pm 
   \frac{\sqrt {-4304\,{R}^{4}{\pi }^{2}{\rho_{{0}}}^{2}-3200\,{R}^{6}{\pi}^{3}{\rho_{{0}}}^{3}-120\,\pi \,{R}^{2}\rho_{{0}}+6400\,{R}^{8}{\pi }^{4}{\rho_{{0}}}^{4}+225}}{32\,{R}^{2}\pi \, \left( 1+ 2\,\pi \,{R}^{2}\rho_{{0}} \right) }\Bigg]. \nonumber
\\ \hspace{-1cm}\label{eq17}
\end{eqnarray}

Again, using Eqs. (\ref{eq11}), (\ref{eq11a}) and (\ref{eq17}) and taking
the value of bag constant as 83 $MeV/{{(fm)}^{3}}$~\cite{Rahaman2014} 
and values of the observed mass as given in Table 1, we get several solutions for $R$. 
However, we examine that among all these solutions only one is physically acceptable 
and also consistent with the Buchdahl condition~\cite{Buchdahl1959}.

\section{Some physical features of the model}\label{sec3}

\subsection{Energy conditions of the system}
To satisfy the energy conditions, i.e., the null energy condition (NEC), weak energy
condition (WEC), strong energy condition (SEC) and dominant energy condition (DEC),  
for an anisotropic fluid sphere composed of strange matter the following inequalities 
have to be hold simultaneously:
\begin{eqnarray}\label{eq21}
&\qquad\hspace{-0.5cm}~NEC:\rho+P_r\geq 0,~\rho+P_t\geq 0, \\ \label{eq22}
&\qquad\hspace{-0.5cm}~WEC: \rho+P_r\geq 0,~\rho\geq 0,~\rho+P_t\geq 0, \\ \label{eq23}
&\qquad\hspace{-0.5cm}~SEC: \rho+P_r\geq 0,~\rho+P_r+2\,P_t\geq 0, \\ \label{eq23a}
&\qquad\hspace{-3cm}~DEC: {\rho}\geq 0, {{\rho}-{P_r}}\geq 0, {{\rho}-{P_t}}\geq 0. 
\end{eqnarray}

\begin{figure}[!htp]\centering
\includegraphics[scale=.3]{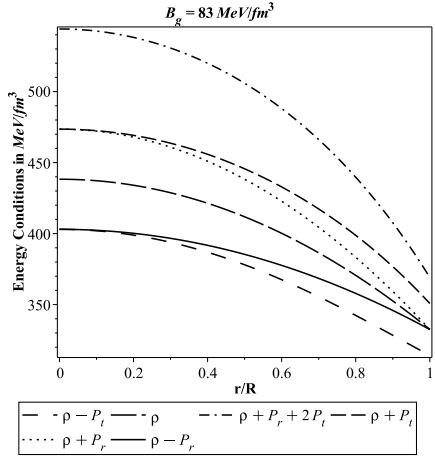}
\caption{Variation of different energy conditions as a function of the fractional radial coordinate $r/R$
	for the strange star $LMC~X-4$. Here value of the bag constant is $B_g= 83~MeV/{{fm}^3}$} \label{Fig6}
\end{figure}

The above energy conditions are drawn in Fig. \ref{Fig6}, which shows that our proposed model for the strange stars
satisfies all the energy conditions.


\subsection{The generalized Tolman-Oppenheimer-Volkoff equations}
According to Tolman~\cite{Tolman1939}, Oppenheimer and Volkoff~\cite{Oppenheimer1939},
the sum of the physical forces should be equal to zero so that the system is subjected to the equilibrium:
\begin{equation}
F_g+F_h+F_a=0, \label{eq20}
\end{equation}
where $F_g$, $F_h$ and $F_a$ are the gravitational, hydrostatic and
anisotropic forces respectively.

The generalized TOV equation~\cite{Leon1993,Varela2010} in its explicit form is
\begin{equation}
-\frac{M_g(\rho +P_r)}{r^{2}} e^{\frac{\lambda- \nu }{2}}
-\frac{d P_r}{dr}+ \frac{2}{r} (P_t-P_r)=0, \label{eq18}
\end{equation}
where $M_g$ is the the effective gravitational mass of the system which is
defined as
\begin{equation}
M_g = \frac{1}{2}r^2e^{\frac{\nu -\lambda}{2}} \nu'. \label{eq19}
\end{equation}

Here for our system ${F_g}=-\frac{1}{2}[\rho+P_{{r}}]\nu'$,
${F_h}=\hspace{-0.1cm}{-{\frac {dP_{{r}} \left( r \right)}{dr}}}$,
and $F_a=2{\frac {(P_{{t}}-P_{{r}})}{r}}$.


\begin{figure}[!htp]\centering
\includegraphics[scale=.3]{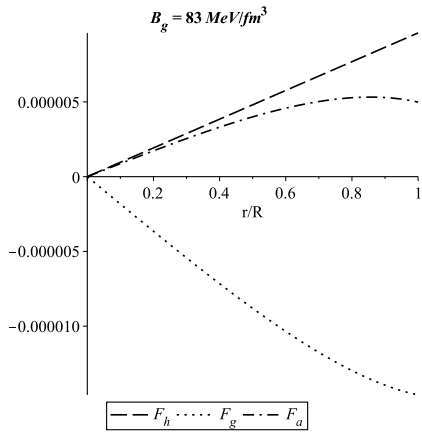}
\caption{Variation of different forces as a function of the fractional radial coordinate $r/R$
	for the strange star $LMC~X-4$. Here value of the bag constant is $B_g= 83~MeV/{{fm}^3}$ } \label{Fig7}
\end{figure}


The nature of above TOV equation is shown in Fig. \ref{Fig7}.
From this it is clear that our proposed model has achieved
equilibrium state under the combined effects of the forces.

\subsection{Herrera cracking concept}
\begin{figure}[!htp]\centering
\includegraphics[scale=.3]{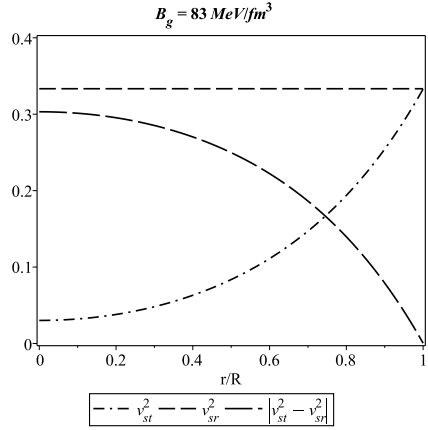}
\caption{Variation of the square of radial $({v_{sr}}^2)$ and tangential $({v_{sr}}^2)$ sound speed and modulus of difference
	of the square of sound speed $(|{v_{st}}^2- {v_{sr}}^2|)$ as a function of the fractional radial coordinate $r/R$ for the   
    strange star $LMC~X-4$. Here value of the bag constant is $B_g= 83~MeV/{{fm}^3}$} \label{Fig8}
\end{figure}

Using the concept of Herrera's cracking~\cite{Herrera1992} we have examined 
the stability of our system. The condition of causality establishes the physical
acceptability of a fluid distribution, which demands $0 \leq {v_{st}}^2 \leq 1$ and
$0 \leq{v_{sr}}^2 \leq 1$. Also, for the stability of the matter distribution the 
condition of Herrera~\cite{Herrera1992} and Andr{\'e}asson~\cite{Andreasson2009} 
is $|{v_{st}}^2- {v_{sr}}^2|\leq 1$ which suggests that
`no cracking' is an essential condition to show a region potentially stable. 
From Fig. \ref{Fig8} we observe that our system satisfies all of these 
conditions and thus provides a stable system.


\subsection{Mass-Radius relation}
Following the work of Buchdahl~\cite{Buchdahl1959} the maximum allowable 
mass-radius ratio for the above proposed anisotropic fluid sphere can be 
calculated. He has proposed that the maximum limit of mass-radius ratio for
static spherically symmetric perfect fluid star should satisfy the
following upper bound $2M/R\leq8/9~(\simeq0.89)$. However, Mak and Harko~\cite{Mak2003}
have given the generalized expression for the same mass-radius ratio.

\begin{figure}[!htp]\centering
\includegraphics[scale=.4]{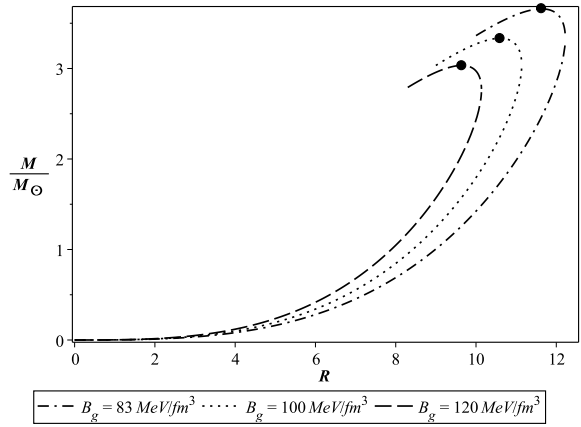}
	\caption{Mass (normalized in solar masses) vs Radius (km) curve for strange stars for the three different values of bag constant.
	Here the solid dots represent maximum-mass star} \label{Fig9}
\end{figure}

\begin{figure}
\centering
\includegraphics[scale=.3]{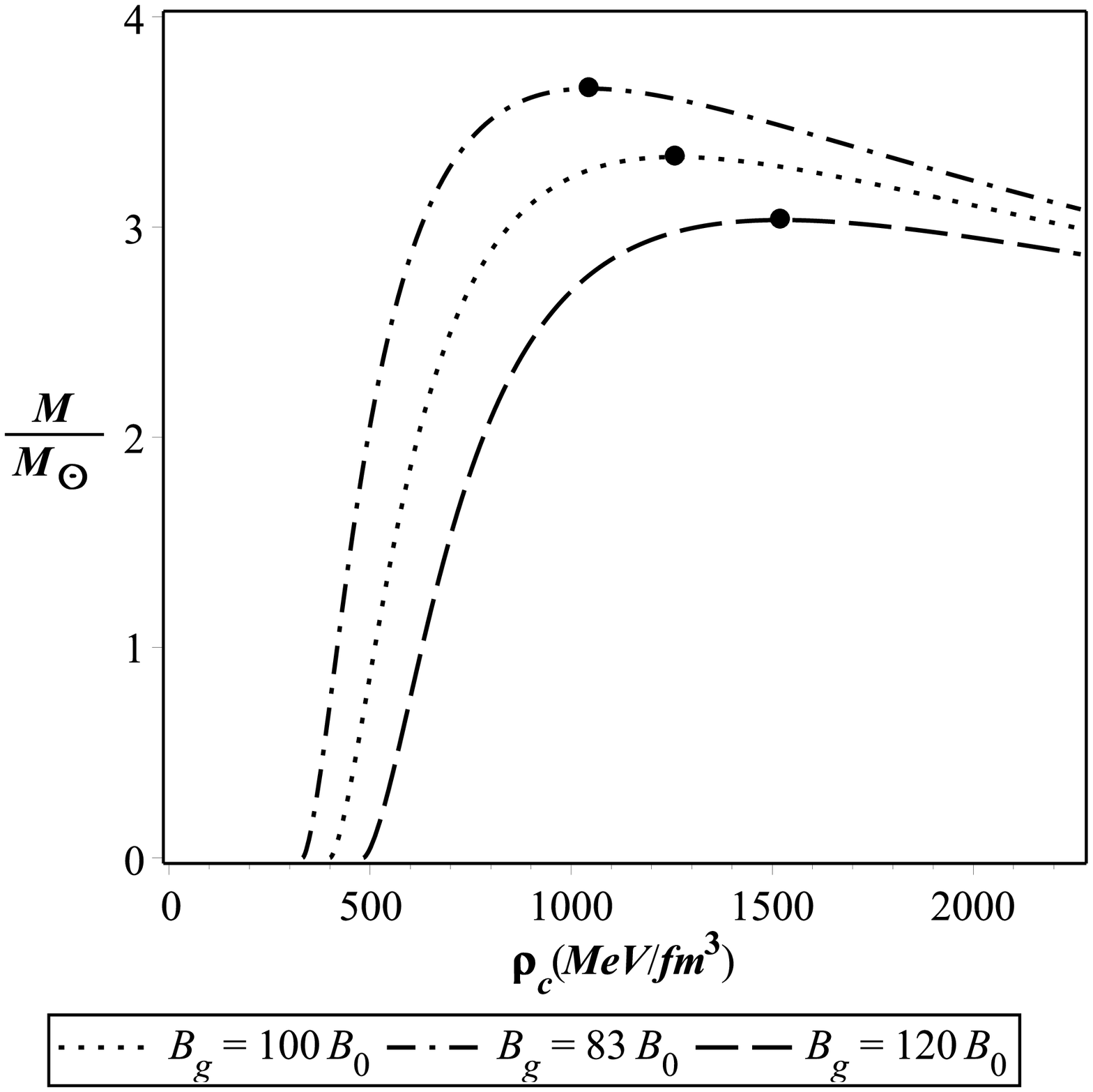}
\includegraphics[scale=.3]{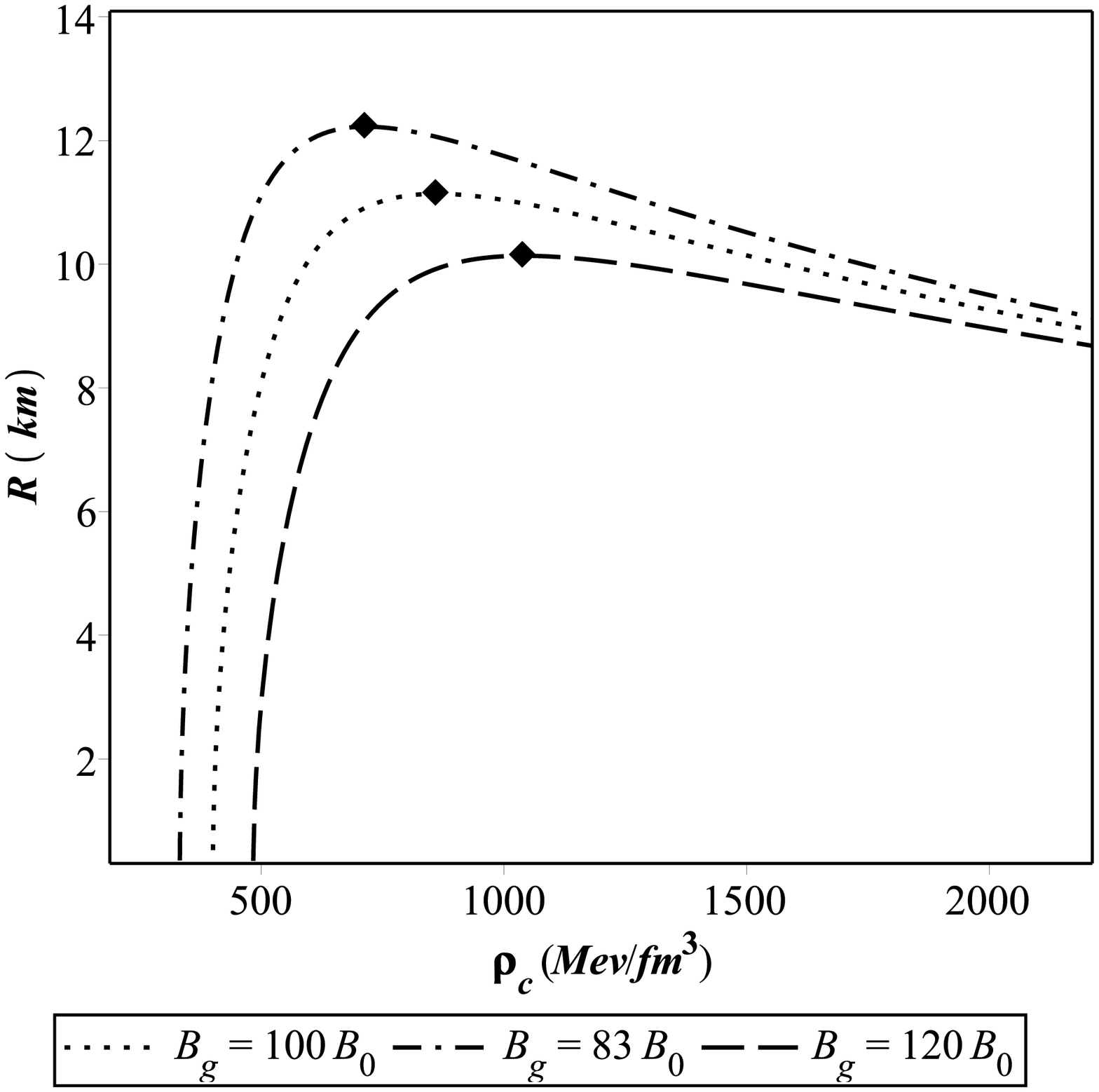}
\caption{Variation of the (i) radius (on the left panel) and (ii) mass (on the right panel) of strange stars as a function of the central density $({\rho}_c)$. Solid diamonds are representing maximum radius and solidcircles are representing maximum mass for the strange stars. Here $B_0$ denotes $1~MeV/{fm^3}$} \label{Fig10}
\end{figure}

The total mass of the anisotropic compact star which is defined
in Eq. (10) can be considered as the maximum mass
$M_{max}=4\,(2\rho_{{c}}+3\rho_0) \pi R^3/15$. From this mass-radius relation
we find out numerical values of different compact star candidates in Table \ref{Table1} and
observe that all the values fall within the acceptable range as specified by
Buchdahl~\cite{Buchdahl1959}. The mass and radius of the strange stars are featured
in Fig. \ref{Fig9} whereas their changes with respect to the central density in Fig. \ref{Fig10}.


\subsection{Surface redshift}
The compactification factor ($u$) for the proposed strange star model can be written as
\begin{equation}
u(r)=\frac{m(r)}{r}={\frac {4\,\pi \,{r}^{2} \left( 5\,\rho_{{c}}{R}^{2}
		-3\,{r}^{2}\rho_{{c}}+3\,{r}^{2}\rho_{{0}} \right) }{15\,{R}^{2}}}.\label{eq24}
\end{equation}

So the corresponding surface redshift ($Z$) as we get from the compactification factor becomes
\begin{equation}
Z = {{\rm e}^{-\frac{\nu\left(R\right)}{2}}}-1={\frac {1}{\sqrt {1-8\,\pi \,{R}^{2} \left( \frac{2}{15}\rho_{{c}}
			+\frac{1}{5}\rho_{{0}} \right) }}}-1.\label{eq25}
\end{equation}

\begin{figure}[!htp]\centering
\includegraphics[scale=.3]{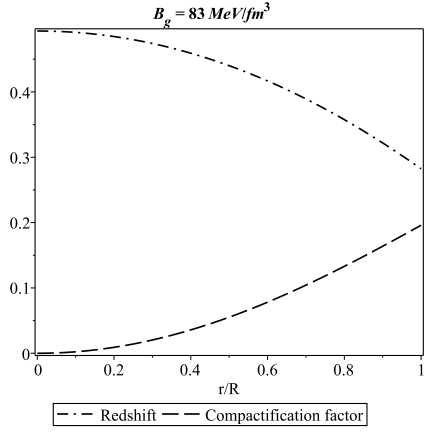}
\caption{Variation of the compactification factor and redshift as a function of the fractional radial coordinate $r/R$
	for the strange star $LMC~X-4$. Here value of the bag constant is $B_g= 83~MeV/{{fm}^3}$} \label{Fig11}
\end{figure}

The variation of the compactification factor and redshift with the fractional radial distance are shown
in the Fig. \ref{Fig11}.


\begin{table*}[!htp]
\centering \caption{Physical parameters as predicted from the proposed model where $1~{{M}_{\odot}}=1.475~km$ for $G=c=1$. The entire data set is valid for ${B_g}=83~MeV/{{(fm)}^{3}}$}
\resizebox{\columnwidth}{!}{
\begin{tabular}{@{}llllllllllllllll@{}}
 \hline Strange & Observed  & Predicted & \hspace{0.5cm}${\rho}_{c}$  & \hspace{0.3cm}${{P_c}}$ & \hspace{-0.1cm}$\frac{2M}{R}$ & $Z$ \\
Stars  & Mass $({{M}_{\odot}})$ & Radius (km) & (${gm/cm}^3$) & \hspace{-0.2cm}(${dyne/cm}^2$)& & \\ 
\hline $PSR~J~1614-2230$ & $1.97 \pm 0.04$ \cite{demorest} & $10.977 \pm 0.06$  & $8.756 \times {10}^{14} $ & $ 8.523 \times {10}^{34} $ &  0.53 &  0.46 
\\ \hline $Vela~X-1$  & $1.77 \pm 0.08$\cite{dey2013} &  $10.654 \pm 0.14$  & $ 8.486 \times 10^{14} $ & $7.603 \times 10^{34}$ & 0.49 & 0.40 \\ 
\hline $PSR~J~1903+327$ & $1.667 \pm 0.021$\cite{dey2013} &  $10.473 \pm 0.037$ & $8.352 \times 10^{14}$ & $7.167 \times 10^{34}$ &  0.47 & 0.37 \\ 
\hline $Cen~X-3$ & $1.49 \pm 0.08$\cite{dey2013} &  $10.136 \pm 0.16$ & $8.082 \times 10^{14}$ & $6.441 \times 10^{34}$ &  0.43 & 0.33 \\ 
\hline $LMC~X-4$ & $1.29 \pm 0.05$\cite{dey2013} &  $9.711 \pm 0.11$ & $7.813 \times 10^{14}$ & $5.654 \times 10^{34}$ & 0.39 & 0.28 \\ 
\hline $4U~1538-52$ & $0.87 \pm 0.07$\cite{dey2013} &  $8.606 \pm 0.215$ & $7.274 \times 10^{14}$ & $4.104 \times 10^{34}$ & 0.30 & 0.19 \\ 
\hline $SMC~X-1$ & $1.04 \pm 0.09$ \cite{dey2013} &  $9.095 \pm 0.243$ & $7.543 \times 10^{14}$ & $4.722 \times 10^{34}$ & 0.34 & 0.23 \\ 
\hline $Her~X-1$ & $0.85 \pm 0.15$\cite{dey2013} &  $8.544 \pm 0.473$ & $7.274 \times 10^{14}$ & $4.031 \times 10^{34}$ & 0.29 & 0.19 \\
\hline $4U~1820-30$ & $1.58 \pm 0.06$\cite{guver2010b} &  $10.312 \pm 0.115$ & $8.217 \times 10^{14}$ & $6.804 \times 10^{34}$ & 0.45 & 0.35 \\ 
\hline $4U~1608-52$ & $1.74 \pm 0.14$\cite{guver2010a} &  $10.603 \pm 0.244$ & $8.486 \times 10^{14}$ & $7.470 \times 10^{34}$ & 0.48 & 0.39 \\ 
\hline $SAX~J1808.4-3658$ & $0.9 \pm 0.3$\cite{dey2013} &  $8.697 \pm 0.924$ & $7.274 \times 10^{14}$ & $4.213 \times 10^{34}$ & 0.31 & 0.20 \\ 
\hline $EXO~1785-248$ & $1.3 \pm 0.2$\cite{ozel2009} &  $9.734 \pm 0.452$ & $7.813 \times 10^{14}$ & $5.678 \times 10^{34}$ & 0.39 & 0.29 \\ 
\hline \label{Table1}
\end{tabular}}
\end{table*}



\begin{table*}[!htp]
\centering \caption{Physical parameters as derived from the proposed model. The entire data set is valid for ${B_g}=83~MeV/{{(fm)}^{3}}$}
\hspace{-0.5cm}\begin{tabular}{@{}llllll@{}}
\hline
Strange & \hspace{0.2cm}$\Delta\left(0\right)$ & \hspace{0.2cm}$\Delta\left(R\right)$ & \hspace{0.2cm}${\Delta}^{\prime\prime}\left(R\right)$ \\
Stars & (in $km^{-2}$) & (in $km^{-2}$) & (in $km^{-2}$) \\
\hline
$PSR~J1614-2230$ & \hspace{0.5cm}$0$ &  $0.000062$ & $-0.0000067$ \\
\hline
$Vela~X-1$ & \hspace{0.5cm}$0$ & $0.000049$ & $-0.0000054$ \\
\hline
$PSR~J1903+327$ & \hspace{0.5cm}$0$ &  $0.000044$ & $-0.0000048$ \\
\hline
$Cen~X-3$ & \hspace{0.5cm}$0$ & $0.000035$ & $-0.0000034$ \\
\hline
$LMC~X-4$ & \hspace{0.5cm}$0$ & $0.000027$ & $-0.0000031$ \\
\hline \label{Table2}
\end{tabular}
\end{table*}


\section{Discussions and Conclusions}\label{sec4}
In this paper employing MIT bag EOS we have presented a new formalism to derive 
exact values of radius of the different strange star candidates
using their observational mass and bag constant (as 83 $MeV/{{fm}^{3}}$) are shown in
Table \ref{Table1}. Also in this table we have computed the values of the central density, 
central pressure, Buchdahl condition and surface redshift for the different strange star candidates. 

From Figs. \ref{Fig1} and \ref{Fig2} it is clear
that our solution is free from geometrical and physical singularities.
Also it can be observed that the metric potentials
${e}^{\lambda(r)}$ and ${e}^{\nu(r)}$ have finite positive values in the range
$0\leq r \leq R$. 

Fig. \ref{Fig3} shows reasonable physical features of the pressures. 
From Eq. (\ref{eq16}) and Fig. \ref{Fig4} one can find that the anisotropic force
$F_a=2\left({P_t}-{P_r}\right)/r$ is positive throughout the system, i.e. ${P_t}>{P_r}$
and hence the direction of the anisotropic force is outward for our system. In Fig. \ref{Fig5} 
variation of the radial (${\omega}_r$) and tangential (${\omega}_t$) EOS parameters 
as a function of the fractional radial coordinate $r/R$. In connection to all plots the strange star $LMC~X-4$
has been taken as the role of the representative. 

The proposed compact stellar model is well consistent with the energy conditions which is clear from 
Fig.~\ref{Fig6}. We find that as sum of all forces ($F_g, F_a$~and~$F_h$) are zero through out the system 
our model is in equilibrium and consistent with the generalized TOV equation (see Fig.~\ref{Fig7}). 
Also from Fig.~\ref{Fig8} we find our model is completely stable under the causality condition and 
Herrera cracking concept. 

According to the well known necessary condition for static stellar model 
mass $M$ should increase with the increasing central density (${\rho}_c$), i.e., $\frac{dM}{d{{\rho}_c}}>0$.   
However, this condition is necessary, but not sufficient one. Again opposite 
condition, i.e., $\frac{dM}{d{{\rho}_c}}<0$ always predicts an unstable model against 
small perturbations. From Figs.~ \ref{Fig9}~and~\ref{Fig10} necessary 
condition $\frac{dM}{d{{\rho}_c}}>0 $ predicts that the stellar system is stable until 
the local maxima, i.e., the maximum mass point. Beyond the maximum mass point mass decreases 
with the increasing density and features instability. As mentioned earlier though the necessary 
condition for stable equilibrium static model is studied and valid for our proposed 
stellar configuration but this is not sufficient.

The equation $\frac{dR}{d{{\rho}_c}}=0$ suggests that the star has maximum value of radius $R_{max}$
for ${{{\rho}_c}|}_{{R}_{max}}$. From equation $\frac{dM}{d{{\rho}_c}}=0$, the star has
maximum value of mass $M_{max}$ for ${{{\rho}_c}|}_{{M}_{max}}$. The model for $83~MeV/{{fm}^3}$ 
yields the numerical values of the maximum mass and radius as $M_{max} = 3.66~{M}_{\odot}$ 
and ${{R}_{max}}$ = 12.225~km respectively which is clear from Fig. \ref{Fig10}.

The detailed analysis of the stability for a non-rotating spherically symmetric anisotropic model 
invites analysis through their radial oscillation but we are keeping this is as future 
project. However, using $M(R)$ curve we are employing a specific criterion which is sufficient 
to determine specific number of unstable normal radial modes~\cite{Haensel2007,Harrison1965}. 
This criterion states that one and only one normal radial mode present at the critical point 
(local maxima or minima) in $M(R)$ curve indicates the change of stability of the system to the 
instability or vice versa. No other points in $M(R)$ curve show the change of stability due to 
the radial oscillation. Also, a mode is unstable only if the $M(R)$ curve bends at the 
critical point counter-clockwise and is stable only if $M(R)$ curve bends clockwise. 
From Fig.~\ref{Fig9} we find that the system is stable up to the critical point, i.e, maximum 
mass point and at that point the $M(R)$ curve bends counter-clockwise making the fundamental 
oscillation mode unstable. That makes the system unstable beyond that critical point as the 
$M(R)$ curve bends to the left counter-clockwise. Finally, applying above technique besides 
the physical tests like equilibrium of forces (generalized TOV equation) and Herrera cracking concept 
we able to predict that the model is stable up to maximum mass point against the small perturbations.

It is shown from the present model that the surface density of the star $LMC~X-4$
is $6.36 \times 10^{14}~gm/cm^3$ which is very high and consistent with the
ultra dense compact stars~\cite{Ruderman1972,Glendenning1997,Herjog2011}. We also have
found out that the redshift of the stars (in Fig. \ref{Fig11} and Table \ref{Table1}) are within the range
0.19-0.46 (for $LMC~X-4$ the maximum mass to radius ratio is $0.40$) 
which is consistent under the constraints $0<Z\leq1$ and
$Z\leq2$ as suggested by the investigators respectively in the following
Refs.~\cite{Kalam2012,Hossein2012,Rahaman2012,Kalam2013,Bhar2015}
and~\cite{Buchdahl1959,Straumann1984,Bohmer2006}. 

Interestingly, in the present study we find an important result that the anisotropy of the compact stars is zero
at the center and then it is increasing through out the interior region of the stars and reach its
maximum value at the surface. Again from Table \ref{Table2} it is clear that the
${\Delta}^{\prime\prime}\left(R\right)$ is negative, which supports mathematically
that the anisotropy is also maximum on the surface of an anisotropic compact star.
As a comparative study, the maximum anisotropy of few quark stars are given in Table \ref{Table2}
(the values are calculated  from our model). The study, therefore, clearly suggests that
the anisotropy is maximum at the surface and it is an inherent characteristics of the singularity free 
anisotropic compact strange stars.

Hence, as a final comment, from the above discussions it can be concluded that the model
proposed in this work seems suitable to study the ultra-dense compact strange stars.

\section*{Acknowledgments}
SR and FR are thankful to the Inter-University Centre for
Astronomy and Astrophysics (IUCAA), Pune, India for providing
Visiting Associateship under which a part of this work was carried
out. SR is also thankful to the authority of The Institute of
Mathematical Sciences (IMSc), Chennai, India for providing all types
of working facility and hospitality under the Associateship scheme. 
A part of this work was performed by DD while he was visiting the 
IUCAA and gratefully acknowledges the hospitality and facilities to continue the work.


\begin{thebibliography}{99}

\bibitem{MH2002} M.K. Mak, T. Harko, Chin. J. Astron. Astrophys. \textbf{2}, 248 (2002)

\bibitem{Ruderman1972} R. Ruderman, Rev. Astron. Astrophys. \textbf{10}, 427 (1972)

\bibitem{BL1974} R.L. Bowers, E.P.T. Liang, Class. Astrophys. J. \textbf{188}, 657 (1974)

\bibitem{HS1997} L. Herrera, N.O. Santos, Phys. Report. \textbf{286}, 53 (1997)

\bibitem{Ivanov2002} B.V. Ivanov, Phys. Rev. D \textbf{65}, 104011 (2002)

\bibitem{SM2003} F.E. Schunck, E.W. Mielke, Class. Quantum Gravit. \textbf{20}, 301 (2003)

\bibitem{MH2003} M.K. Mak, T. Harko, Proc. R. Soc. A \textbf{459}, 393 (2003)

\bibitem{Usov2004} V.V. Usov, Phys. Rev. D \textbf{70}, 067301 (2004)

\bibitem{Varela2010} V. Varela, F. Rahaman, S. Ray, K. Chakraborty, M. Kalam, Phys. Rev. D \textbf{82}, 044052
(2010)

\bibitem{Rahaman2010} F. Rahaman, S. Ray, A.K. Jafry, K. Chakraborty, Phys. Rev. D \textbf{82}, 104055 (2010)

\bibitem{Rahaman2011} F. Rahaman, P.K.F. Kuhfittig, M. Kalam, A.A. Usmani, S. Ray, Class. Quantum Gravit.
\textbf{28}, 155021 (2011)

\bibitem{Rahaman2012} F. Rahaman, R. Maulick , A.K. Yadav, S. Ray, R. Sharma, Gen. Relativ. Gravit. \textbf{44}, 107
(2012)

\bibitem{Kalam2012} M. Kalam, F. Rahaman, S. Ray, Sk.M. Hossein, I. Karar, J. Naskar, Eur. Phys. J.
C \textbf{72}, 2248 (2012)

\bibitem{Deb2015} D. Deb, S.R. Chowdhury, S. Ray, F. Rahaman, arXiv:1509.00401 [gr-qc]
	
\bibitem{Shee2016} D. Shee, F. Rahaman, B.K. Guha, S. Ray, Astrophys. Space Sci. \textbf{361}, 167 (2016)

\bibitem{Maurya2016} S.K. Maurya, Y.K. Gupta, S. Ray, D. Deb, Eur. Phys. J. C \textbf{76}, 693 (2016)

\bibitem{Maurya2017} S.K. Maurya, D. Deb, S. Ray, P.K.F. Kuhfittig, eprint arXiv:1703.08436

\bibitem{Dev2003} K. Dev, M. Gleiser, General Relativity and Gravitation \textbf{35}, 1435 (2003) 

\bibitem{Kalam2005} M. Kalam, A. Usmani, F. Rahman, I. Karar, R. Sharma, Int. J. Theor. Phys. \textbf{52}, 3319 (2013)

\bibitem{Horvat2011} D. Horvat, S.Ilijic, A. Marunovic, Class. Quantum Grav. 28 (2011) 025009 

\bibitem{Itoh1970} N. Itoh, Prog. Theor. Phys. \textbf{44}, 291 (1970)

\bibitem{Bodmer1971} A.R. Bodmer, Phys. Rev. D \textbf{4}, 1601 (1971)

\bibitem{Cheng1998} K.S. Cheng, Z.G. Dai, T. Lu, Int. J. Mod. Phys. D \textbf{7}, 139 (1998)

\bibitem{Alford2001} M. Alford, Ann. Rev. Nucl. Part. Sci. \textbf{51}, 131 (2001)

\bibitem{Alcock1986} C. Alcock, E. Farhi, A. Olinto, Astrophys. J. \textbf{310}, 261 (1986)

\bibitem{Haensel1986} P. Haensel, J. L. Zdunik, R. Schaefer, Astron. Astrophys.  \textbf{160}, 121 (1986)

\bibitem{Weber2005} F. Weber, Prog. Part. Nucl. Phys. \textbf{54}, 193 (2005)

\bibitem{Perez2010} M. A. Perez-Garcia, J. Silk, J.R. Stone, Phys. Rev. Lett.  \textbf{105}, 141101 (2010)

\bibitem{Rodrigues2011} H. Rodrigues, S.B. Duarte, J.C.T. de Oliveira, Astrophys. J. \textbf{730}, 31 (2011)

\bibitem{Bordbar2011} G.H. Bordbar, A.R. Peivand, Res. Astron. Astrophys. \textbf{11}, 851 (2011)

\bibitem{1} M. Brilenkov, M. Eingorn, L. Jenkovszky, A. Zhuk, JCAP \textbf{08} 002 (2013)

\bibitem{2} N.R. Panda, K.K. Mohanta, P.K. Sahu, J. Physics: Conference Series \textbf{599}, 012036 (2015)

\bibitem{3} A.A. Isayev, Phys. Rev. C \textbf{91}, 015208 (2015)

\bibitem{4} S.D. Maharaj, J.M. Sunzu, S. Ray, Eur. Phys. J.Plus. \textbf{129}, 3 (2014)

\bibitem{5} L.Paulucci, J.E.Horvath, Physics Letters B \textbf{733}, 164 (2014)

\bibitem{6}  G. Abbas, S. Qaisar, A. Jawad, Astrophys. Space Sci. \textbf{359}, 57 (2015)

\bibitem{7} J.D.V. Arba{\~n}il, M. Malheiro, JCAP \textbf{11}, 012 (2016)

\bibitem{8} G. Lugones, J.D.V. Arba{\~n}il, Phys. Rev. D \textbf{95}, 064022 (2017)

\bibitem{Rahaman2014} F. Rahaman, K. Chakraborty, P.K.F. Kuhfittig, G.C. Shit, M. Rahman, Eur. Phys. J. C \textbf{74}, 3126 (2014)

\bibitem{Chodos1974} A. Chodos, R.L. Jaffe, K. Johnson, C.B. Thorn, V.F. Weisskopf, Phys. Rev. D \textbf{9}, 3471 (1974)

\bibitem{Farhi1984} E. Farhi, R.L. Jaffe, Phys. Rev. D \textbf{30}, 2379 (1984)

\bibitem{Witten1984} E. Witten, Phys. Rev. D \textbf{30}, 272 (1984)

\bibitem{Buchdahl1959} H.A. Buchdahl, Phys. Rev. D \textbf{116}, 1027 (1959)

\bibitem{Tolman1939} R.C. Tolman, Phys. Rev. \textbf{55}, 364 (1939)

\bibitem{Oppenheimer1939} J.R. Oppenheimer, G.M. Volkoff, Phys. Rev. \textbf{55}, 374 (1939)

\bibitem{Leon1993} J. Ponce de Le{\'o}n, Gen. Relativ. Gravit. \textbf{25}, 25 (1993)

\bibitem{Herrera1992} L. Herrera, Phys. Lett. A \textbf{165}, 206 (1992)

\bibitem{Andreasson2009} H. Andr{\'e}asson, Commun. Math. Phys. \textbf{288}, 715 (2009)

\bibitem{Mak2003} M.K. Mak, T. Harko, Proc. R. Soc. A \textbf{459}, 393 (2003)

\bibitem{demorest} P.B. Demorest, T. Pennucci, S.M. Ransom, M.S.E. Roberts, J.W.T. Hessels, Nature {\bf 467}, 1081 (2010)

\bibitem{dey2013} T. Gangopadhyay, S. Ray, X.-D. Li, J. Dey, M. Dey, Mon. Not. R. Astron. Soc. {\bf 431}, 3216 (2013)

\bibitem{guver2010b} T. G{\"u}ver, P. Wroblewski, L. Camarota, F. {\"O}zel, Astrophys J. {\bf 712}, 964 (2010)

\bibitem{guver2010a} T. G{\"u}ver, P. Wroblewski, L. Camarota, F. {\"O}zel, Astrophys J. {\bf 719}, 1807 (2010)

\bibitem{ozel2009} F. {\"O}zel, T. G{\"u}ver, D. Psaltis, Astrophys J. {\bf 693}, 1775 (2009)

\bibitem{Haensel2007} P. Haensel, A.Y. Potekhin, D.G. Yakovlev, \textit{Neutron Stars 1: Equation of State and Structure} (Springer Verlag, New York, 2007)

\bibitem{Harrison1965} B.K. Harrison, K.S. Thorne, M. Wakano, J.A. Wheeler, \textit{Gravitation Theory and Gravitational Collapse} (University of Chicago Press, Chicago, 1965)

\bibitem{Glendenning1997} N.K. Glendenning, Compact Stars: Nuclear Physics, Particle Physics and General Relativity (Springer, New York, 1997)

\bibitem{Herjog2011} M. Herzog, F.K.R{\"o}pke, Phys. Rev. D \textbf{84}, 083002 (2011)

\bibitem{Hossein2012} S.M. Hossein, F. Rahaman, J. Naskar, M. Kalam, S. Ray, Int. J. Mod. Phys. D \textbf{21}, 21 (2012)

\bibitem{Kalam2013} M. Kalam, A.A. Usmani, F. Rahaman, S.M. Hossein, I. Karar, R. Sharma, Int. J. Theor. Phys. \textbf{52}, 3319 (2013)

\bibitem{Bhar2015} P. Bhar, F. Rahaman, S. Ray, V. Chatterjee, Eur. Phys. J. C \textbf{75}, 190 (2015)

\bibitem{Straumann1984} N. Straumann, General Relativity and Relativistic Astrophysics (Springer, Berlin, 1984)

\bibitem{Bohmer2006} C.G. B{\"o}hmer, T. Harko, Class. Quantum Gravit. \textbf{23}, 6479 (2006)

\end{thebibliography}
\end{document}